\documentclass[a4paper,11pt]{article}

\usepackage{acronym}
\usepackage{jinstpub}
\usepackage{gensymb}
\usepackage{xcolor}
\usepackage{caption}
\usepackage{subcaption}
\usepackage{amsmath}
\usepackage{indentfirst}
\usepackage{booktabs}

\usepackage{lineno}
\linenumbers

\title{Gaseous argon time projection chamber with electroluminescence enhanced optical readout }

\author[a]{R.M. Amarinei,\footnote{Corresponding author.}}
\author[a]{F. S\'anchez,}
\author[a]{S. Bordoni,}
\author[b]{T. Lux,}
\author[a]{L. Giannessi, }
\author[a]{E. Roe,}
\author[c]{E. Radicioni}

\affiliation[a]{Département de physique nucléaire et corpusculaire - The particle physics department (DPNC), University of Geneva,  1205 Gen\`eve, Switzerland}
\affiliation[b]{Institut de Física d’Altes Energies (IFAE)—The Barcelona Institute of Science and Technology (BIST), Campus UAB, E-08193 Bellaterra
(Barcelona), Spain}
\affiliation[c]{Dipartimento Interuniversitario di Fisica, INFN Sezione di Bari and Universita` e Politecnico di Bari (Bari), Italy}

\emailAdd{robert-mihai.amarinei@unige.ch}

\abstract{
Systematic uncertainties in accelerator oscillation neutrino experiments arise mostly from nuclear models describing neutrino-nucleus interactions. To mitigate these uncertainties, we can study neutrino-nuclei interactions with detectors possessing enhanced hadron detection capabilities at energies below the nuclear Fermi level. Gaseous detectors not only lower the particle detection threshold but also enable the investigation of nuclear effects on various nuclei by allowing for changes in the gas composition. This approach provides valuable insights into the modelling of neutrino-nucleus interactions and significantly reduces associated uncertainties.
Here, we discuss the design and first operation of a gaseous argon time projection chamber optically read. The detector operates at atmospheric pressure and features a single stage of electron amplification based on a thick GEM. Here, photons are produced with wavelengths in the vacuum ultraviolet regime. In an optical detector the primary constraint is the light yield. This study explores the possibility of increasing the light yield by applying a low electric field downstream of the ThGEM. In this region, called the electroluminescence gap, electrons propagate and excite the argon atoms, leading to the subsequent emission of photons. This process occurs without any further electron amplification, and it is demonstrated that the total light yield increases up to three times by applying moderate electric fields of the order of 3~kV/cm. Finally, an indirect method is discussed for determining the photon yield per charge gain of a ThGEM, giving a value of 18.3 photons detected per secondary electron. }


\arxivnumber{1234.56789} 

\nolinenumbers
\begin{document}\maketitle

\flushbottom


\section{Introduction}




One of the primary sources of system uncertainties in current accelerator neutrino experiments is the too-high particle momentum detection threshold. In modern experiments like T2K~\cite{T2K}, the particle detection threshold  is approximately 450~MeV/c. This poses a problem because neutrino interactions can produce hadrons with momenta below this threshold, resulting in the incorrect classification of the neutrino interaction.

To illustrate this issue, let's consider the interaction of a neutrino with energies ranging from several hundred MeV to a few GeV via charged current exchange with the detector medium. During this process, a lepton is generated and can be easily detected. However, in a 2p2h neutrino interaction, for instance, two protons are emitted from the nucleus in the final state, but most probable, only one of them carries enough energy to surpass the detection threshold~\cite{Federico_god}. Consequently, these low-energy protons remain undetected, leading to systematic errors in the final measurements.

To enable the detection of low-energy protons and therefore mitigate the systematic uncertainties, it is important to develop a detector with a lower momentum detection threshold. In this paper, we explore the potential of a gaseous optical Time Projection Chamber (TPC) based on pure argon gas, operated at atmospheric pressure. A TPC consists of a gas filled volume where a homogeneous and uniform electric field is applied, enclosed by a field cage, a cathode, and a segmented anode. When a particle crosses the TPC active volume it produces ion-electron pairs, which start to \\propagate towards the anode. A TPC allows for 3D tracking by reconstructing the projected electron track on the anode plane and using the time information from the primary electrons reaching the anode from the ionisation point. 

Using the TPC with a gaseous material ensures that even a low-energy hadron can propagate for long enough to reconstruct its track. A problem with the gaseous material approach is that a minimum ionising particle only produces around 90 electrons (primary electrons) per cm in argon, not enough to be detected directly. For this reason, an amplification stage is installed directly in front of the anode. The amplification stage is based on a Thick Gaseous Electron Multiplier (ThGEM), consisting of a region with a very high electric field. An electron released in the drift volume entering the ThGEM undergoes a charge avalanche process in which up to a few thousand electron-ion pairs are created. The signal (charge) amplified in this way is sufficient to be detected with cost-effective electronics, and thus, the track of the incident particle is reconstructed.  

Another approach to reconstructing the incident particle's track is to detect the photons produced during the electron avalanche. A TPC in which the readout is done using photons is called an optical TPC. This technology is not new, but it only became interesting recently with the evolution of fast and highly pixelated photon sensors~\cite{TimePix4}. The advantages of using an optical readout include that the photon sensors are decoupled from the TPC chamber, thus avoiding sensor degradation and gas contamination over time. Another advantage is that large detector areas can be equipped with a relatively small number of sensors if appropriate lens systems are used. 

However, optical TPCs are limited by the minimal amount of photons arriving at the detection plane. The light yield of gaseous argon TPCs (based on GEMs and ThGEMs) is studied in detail in~\cite{Monteiro:2008zz}, \cite{MONTEIRO2012_ThGEM_GEM},  and \cite{ArgonLight:BUZULUTSKOV}. Because photons are emitted uniformly in 4$\pi$,  a significant fraction does not reach the detection plane. Furthermore, the photons emitted through argon de-excitations have wavelengths in the Vacuum Ultraviolet (VUV) spectrum. To match the sensitivity of standard photon detectors, the wavelength has to be increased to the visible spectrum. This is done using a wavelength shifter, decreasing further the photon number which can be effectively detected. 

It becomes clear that maximising the photon yield is of primary importance when designing an optical TPC. Our work touches on this point precisely: we investigate the possibility of enhancing the light yield by placing an additional small electric field downstream of the ThGEM, where the electrons produced in the avalanche multiplication can drift further and produce additional photons. To our knowledge, only one similar work has been done in the past (CYGNO collaboration~\cite{CYGNO_2020}); however, that study was done with an optical TPC based on a He/CF$_4$ gas mixture and several GEMs for photon production.

The present paper  is structured as follows: in section \ref{sec:ExperimentalSetup} a complete description of the experimental setup is given, continued in section \ref{Sec:LightProduction} with a description of the light production within the detector.  Section \ref{sec:DataProc} describes the data pre-processing and the calibration of the photon detector, section \ref{sec:Results} focuses on the results, and finally, the paper conclusions are drawn in section \ref{sec:Concl}.

\section{Experimental setup}
\label{sec:ExperimentalSetup}
\begin{figure}[htb]
    \centering
    \includegraphics[width = 0.8\textwidth]{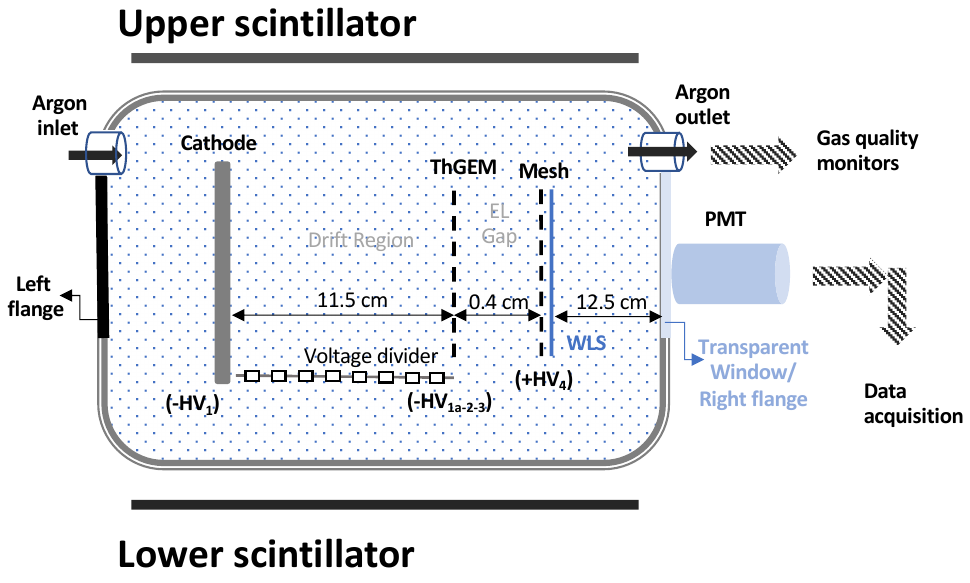}
    \caption{ Schematic view of the experimental setup. The depicted components and layout are not to scale. Distance from cathode to ThGEM: 11.5~cm. Distance from ThGEM to WLS/mesh: 0.44~cm. Distance from WLS/mesh to window: 12.5~cm. The PMT faces the TPC window less than 1~cm away from it. }
    \label{fig:TPCScheme}
\end{figure}

The TPC used in this work (figure~\ref{fig:TPCScheme}) is placed inside a $30$~l cylindrical stainless-steel barrel, closed on each side with a removable flange. The flange on the left side of the cathode contains 6 feed-throughs: 5 are used to power the components inside the volume and  one for the gas inlet. The flange on the right side hosts the gas outlet and a fused silica transparent window. Inside the tank, a robust peek-3000 frame is installed in order to accommodate all the detector parts. Peek-3000 is a plastic typically used in high purity environments due to its excellent outgassing properties~\cite{NASA}, and moreover, having a high dielectric strength ($190$~kV/cm~\cite{emc}), it proves suitable for insulation between the different electric fields applied within the chamber.   

The drift region of the TPC is defined by three components: a cathode, a drift field cage, and a ThGEM which acts as anode. The cathode is a $10$~cm radius copper disk fitted with holes for a better flow of gas. Tested for stability at up to $-30$~kV, during the present measurements, the cathode is kept at $-3500$~V with respect to ground. To obtain a uniform electric field, the cathode is coupled to the drift cage which consists of 16 equally spaced copper strips with $50$~M$\Omega$ resistors between each of them. The Last Field Strip (LFS) is kept at a voltage of $-1480$~V, thus giving a drop of $126.25$~V between each strip. The LFS is placed 10~cm from the cathode and therefore gives rise to a drift electric field of 200~V/cm\footnote{At this drift field strength, the expected drift velocity for an electron is $\sim$~0.3~cm/$\mathrm{\micro}$s.}. To close the drift field, the anode (ThGEM) is placed 1.5~cm from the LFS. For simplicity, the ThGEM is completely disconnected from the cathode and LFS, being kept at a constant voltage of -1400~V through a power supply of its own. This closes the drift field with an electric field between the LFS and the ThGEM of 53~V/cm. Even though this value is smaller than the nominal drift field strength of 200~V/cm, it is expected that this difference is not going to impact the conclusion of this paper; the difference comes from a measurement imprecision, which was found post data taking.

The  ThGEM has a typical square geometry with an active area of 100~cm$^2$, prolonged with two 4~cm conductive strips in order to facilitate the voltage application. The ThGEM holes (rim-less) are made through mechanical drilling, having a radius of $200$~$\mathrm{\micro}$m and being placed in a hexagonal pattern with 800~$\mathrm{\micro}$m pitch. The two conductive sides (upstream/downstream ThGEM) are made of $0.025$~mm thin copper, separated by a 1~mm layer of insulating material (PCB material: EM-370(5)/EM-37B(D)). On the plate facing the cathode, a voltage of $-1400$~V is applied, while the downstream plate is grounded, giving a field strength of $14$~kV/cm inside the ThGEM holes. This field is intense enough to produce a moderate electron cascade during which both electrons and photons are produced; the photons are detected by later stages of the detector ensemble. A detailed description of the light production and detection is presented in section~\ref{Sec:LightProduction}. 

Further down, a conductive mesh is placed at a distance of $0.44$~cm from the ThGEM. The mesh has a hexagonal structure with a side of 6.5~mm and a thickness of 0.5~mm. The region between the downstream ThGEM and the mesh is referred to in the following as the electroluminescence (EL) gap. The light produced at this stage is presented  in detail in section \ref{sec:Results}. A summary of the main elements of the experimental setup, as well as the electric fields applied in the various regions, is given in table~\ref{tab:OperatingConditions}.

It is known that photons produced by argon gas span a broad wavelength spectrum, with a large component in the VUV regime~\cite{Santorelli:2020fxn}. A Hamamatsu R6427  Photomultiplier Tube (PMT) is used to detect these photons, and, to match its sensitivity, a Polyethylene Naphthalate (PEN) film, provided by GoodFellow, is used as a Wavelength Shifter (WLS). This consists of a $0.125$~mm film, which has a wavelength shifting efficiency of $25$\% for incident photons with $128$~nm wavelength, re-emitting them at 420~nm~\cite{PEN:Kuzniak}. Using a peek-3000 frame, the WLS is pressed against the mesh with the aim to reduce the distance between the two as much as possible. This has a two-fold importance for the detector. Firstly, the PEN is a non-conductive material, so any possible charge deposition is dispersed by the mesh. Secondly, the semi-rigid structure of the mesh is used as support for the flexible film, ensuring that this lays in a plane perpendicular to the drift field direction.

The PMT used to collect the light from the TPC is a Hamamatsu R6427~\cite{hamamatsu} coupled to a Hamamatsu E2624-14 base. This is placed outside of the anode flange, at very few mm distance to the silica window. Its peak quantum efficiency wavelength matches exactly the emission wavelength of the PEN re-emitted photons (420~nm). In order to increase the sensitivity to single photon detection, the PMT is biased to a high voltage of $-1650$~V. 

The TPC is triggered by an external cosmic ray telescope, built of plastic scintillator bars which are coupled to the same model of PMTs as mentioned above. The two bars are placed parallel to the drift volume, slightly extending beyond the limits of the active TPC volume. This negatively impacts the geometrical efficiency, which is estimated to be $\sim$50\%, meaning that only half of the cosmic rays crossing the two scintillator bars pass through the drift volume. If signals from the two scintillator bars are recorded within a 3.5~ns window, the readout of the TPC PMT is triggered, storing a 100~$\mathrm{\micro}$s waveform. The logic coincidence between the two scintillator bars is done through a dedicated LeCroy~465 coincidence unit, and the data acquisition and monitoring is based on a LeCroy WaveJet touch 354 oscilloscope with up to 1GHz sampling.

During operation, the detector is continuously circulated with pure argon gas, kept at a constant pressure of 1~bar. The gas system is conceptually simple: the gas flux coming from an argon tank is controlled by several instruments upstream the TPC, followed by a suite of purity sensors and pressure-controlling devices downstream.  The gas flow is set and maintained by a Bronkhorst EL-Flow device with a range of 0 to 30~l/h, followed by a Hoke (reference  LR6032-6Z-BH) safety valve which prevents pressures higher than 2~bar from building up within the system. Past these two elements, the gas enters and exits the TPC at a fixed flow of 15~ln/h. Downstream of the TPC, a Bronkhorst EL-Press Back Pressure Regulator (BPR) is used to maintain a fixed pressure within the detector, followed by a series of 3 devices which monitor the oxygen, water contamination, and temperature of the gas. Their manufacturers and models are: Oxi.Iq (Oxygen transmitter 1-1-2-0-0-0), Vaisala (H$_2$O sampling cell DMT242SC2), and Jumo (PT100 probe) respectively. All of the gas system elements described earlier are operated remotely, and the detector is monitored constantly in order to ensure it is operating at the nominal conditions summarised in table \ref{tab:OperatingConditions}. The N$_2$ contamination is not measured with the current setup. However, the gas supplier claims a similar N$_2$ contamination to O$_2$. Therefore, it is assumed that all the relevant pollutants are kept well below the level in which they could negatively impact the measurements.

\begin{table}[htb]{
\begin{tabular}{|l|c|c|}\hline
 \textbf{Component}  	& \textbf{Value}	&  \textbf{Comment}   \\
  	\hline
 Cathode LFS distance  	&   $100$~mm  &  \\
 LFS ThGEM distance& $15$~mm&\\
EL gap distance    &    $4$~mm  &  \\
 Distance PEN-window &    $125$~mm  &  \\
 Drift volume diameter & $200$~mm &  \\
 Number of field strips &   $16$  & Copper-coated Kapton foils  \\
 \hline
  ThGEM thickness &   $1$~mm  &     \\
  ThGEM hole diameter &   $400$~$\mathrm{\micro}$m   & \\    
    ThGEM hole pitch &  $800$~$\mathrm{\micro}$m   & \\  
\hline
 Mesh thickness  & $0.5$~mm & Aluminium\\
Mesh wire thickness  & $0.5$~mm & \\
 Mesh optical transparency & $78$\%  & Photons perpendicular to mesh   \\
 \hline 
 Cathode voltage (-HV$_1$)  &  $-3500$~V & Copper coated G10 \\
 Last field strip voltage (-HV$_{\text{1a}}$)  &  $-1480$~V &  \\
 ThGEM upstream voltage (-HV$_2$)&  $-1400$~V &  \\
 ThGEM downstream voltage (-HV$_3$) &   $0$~V &  \\
 Nominal Mesh voltage (+HV$_4$)& 0 - 1400~V& \\
 Cathode $\rightarrow$ LFS field & $200$~V/cm  & \\
 LFS $\rightarrow$ ThGEM up field & $53$~V/cm  & \\
 \hline 
  Optical window thickness &  $8.3$~mm   &  \\ 
  Optical window diameter &  $100$~mm   & Fused silica \\ 
  Optical window transparency range & $200-2200$~nm     &  \\ 
Optical window index of refraction & $1.458$ &  \\
 \hline 
 Wavelength shifting thickness &   $0.125$~mm        & Polyethylene Naphthalate   \\
  Wavelength shifting conversion efficiency &   $0.25$        & Biaxially Oriented (GoodFellow) \\
 \hline
    PMT  active area diameter & $25$~mm &  \\
    PMT  distance to window & few~mm &  \\
    PMT spectral response & $300-650$~nm & Hamamatsu R6427 \\
    PMT  operational voltage & $1650$~V &  \\     
  	\hline
 Gas pressure       &  $1$~bar  & Fixed by flow controller \\
 Oxygen contamination &  $3$~ppm & \\
 Water contamination & $50$~ppm & \\
 Gas temperature & $20$~\degree C&\\
 Gas replacement rate & $2$~volumes/hour & \\
 \hline
\end{tabular}
\caption{Summary of the TPC geometry and nominal operating parameters throughout the entire data taking.}
\label{Tab:TPCSettings}
\label{tab:OperatingConditions}
}
\end{table}

\section{Light production mechanisms in argon}
\label{Sec:LightProduction}

Light produced in argon is extensively used in current experiments such as ICARUS~\cite{ICARUS:2020wmd}, DUNE~\cite{DUNE:2021hwx,Belver:2018erf}, MICROBOONE~\cite{Conrad:2015xta}, SBND~\cite{Szelc:2016rjm}, and DARKSide~\cite{DarkSide:2018kuk}. Despite its popularity, the light production mechanism of argon is not fully understood, and it is still an intensely studied field of physics. Argon emits light in a broad wavelength spectrum, covering the Vacuum Ultra-Violet (VUV), Ultra-Violet (UV), Visible Light (VL) and Infra-Red (IR) regimes.

One of the most studied light emissions is due to excimer decay, producing photons in the VUV regime with a peak at $128$~nm.

\begin{equation}\label{eq:ExcimerDecay}
     \text{Ar}_2^* \rightarrow 2\text{Ar} +\text{h}\nu  \text{ (128~nm photons)}
\end{equation}

A less understood emission of argon is the so called third-continuum. This consists of the radiative transition between two molecular ion levels, generally populated by three body collisions involving an atomic excited state. It was shown that this emission occurs with a 10 times lower probability than the emission due to the excimer decay (equation~\ref{eq:ExcimerDecay}) and also that it does not occur in the solid or liquid state of argon~\cite{Klein_1981}.   

\begin{equation}\label{eq:UVTransition}
     \text{Ar}^{2+}_2 \rightarrow 2\text{Ar} +\text{h}\nu  \text{ (160 - 290~nm)}
\end{equation}

For a long time it was believed that argon emits light mostly in the VUV spectrum. Recently, however, measurements showed that there is a significant light component with wavelengths spanning from 100~nm to 1000~nm. This was explained through the theory of Neutral Bremsstrahlung (NB), according to which slow electrons scatter elastically or inelastically off neutral argon atoms, yielding light compatible with the above-mentioned values. In the same study, it was also shown that the main component of NB comes from elastic collisions of electrons with energies between $1-10$~eV \cite{ArgonLight:BUZULUTSKOV}. 

\begin{equation}\label{eq:Neutralbrem}
    \text{e}^- + \text{Ar} \rightarrow \text{e}^- + \text{Ar} + \text{h}\nu
\end{equation}
\vspace{-0.7cm}
\begin{equation}
    \text{e}^- + \text{Ar} \rightarrow \text{e}^- + \text{Ar}^* + \text{h}\nu
\end{equation}

All of these reactions are allowed to happen in the present detector, but they will quantitatively differ in different regions of it. In the TPC, light is produced mainly in the ThGEM holes and in the so-called EL gap; a detailed description of light production in the different detector regions follows.

As the ionisation electrons undergo avalanche multiplication in the ThGEM, a copious amount of argon atoms are excited and ionised. In this phase, all the aforementioned light production mechanisms are possible; photons are produced towards the exit of the holes~\cite{Lux_2019} and may be detected by the PMT. Further, by applying a voltage on the mesh placed downstream of the ThGEM, electrons are extracted from the holes and start to propagate across the EL gap. By tuning the electric field in this region, the drifting electrons are given enough energy to excite the argon atoms but not to ionise them. We expect the excimer decay (equation~\ref{eq:ExcimerDecay}) to be the most relevant in the EL gap (where the reduced electric field does not exceed 13 Td) with some small contribution from NB (equation~\ref{eq:Neutralbrem}) in the low reduced field region (below $\sim$4~Td). In this paper we refer to these two mechanisms, and in general to the light produced in the EL gap, as EL light.
On the other hand, we expect the less frequent third-continuum reaction (equation~\ref{eq:UVTransition}) to be negligible in this region, since the average electron energy is not large enough to ionise the argon atoms, thus the creation of the ion-molecule $\text{Ar}^{2+}_2$ becomes irrelevant. However, this mechanism might give a small contribution in the light produced in the ThGEM, where the high electric field and electron density might allow the creation of the unstable $\text{Ar}^{2+}_2$ state. 
It is worth to be mentioned here that for each of the aforementioned light production mechanisms, photons are produced uniformly in the whole 4$\pi$ solid angle, thus only a small fraction of them will be detected by the PMT.

\begin{figure}[htb]
    \centering
    \includegraphics[width = 0.8\textwidth]{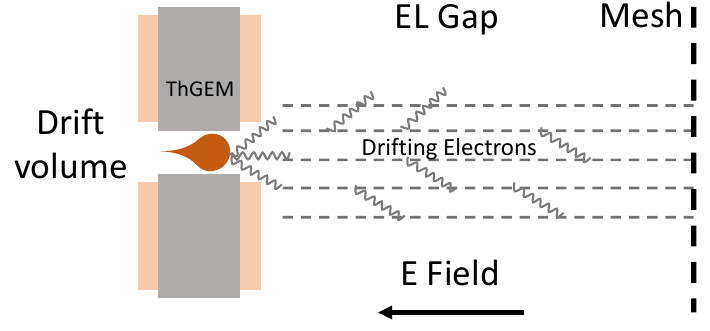}
    \caption{Light production in the ThGEM and EL gap: as an electron enters a ThGEM hole it undergoes avalanche multiplication, where, at the same time, it excites the argon atoms, causing the emission of photons (wavy lines). The E field then extracts the electrons from the holes and they drift towards the mesh. On their path they cause the uniform emission of light via the electroluminescence process. Scheme is not drawn to scale. } 
    \label{fig:ThGEM_Hole}
\end{figure}

Figure~\ref{fig:ThGEM_Hole} shows a schematic depiction of the production of light in the detector.

\section{Data pre-processing}

\label{sec:DataProc}

\subsection{Processing TPC wave-forms  }

\begin{figure}[htb]
        \centering
         \includegraphics[width=0.7\textwidth]{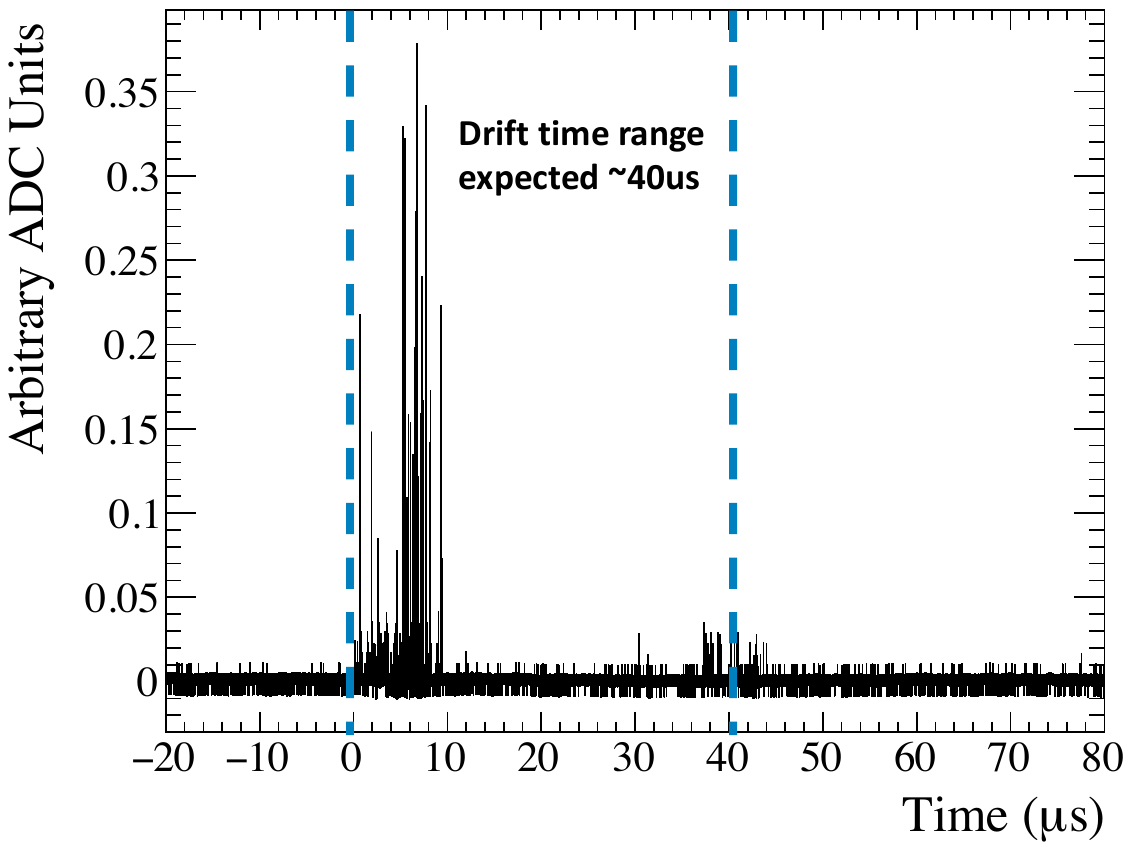}
         \caption{Typical waveform showing a particle crossing the drift volume in the ThGEM proximity. The blue dashed lines represent the nominal drift time for an ionisation electron produced in proximity of the ThGEM (t$\simeq$0~$\mathrm{\micro}$s) and in the proximity of the cathode (t$\simeq$40~$\mathrm{\micro}$s).}
         \label{fig:WF_Exp}
\end{figure}

Figure~\ref{fig:WF_Exp} shows a typical TPC signal event waveform recorded in coincidence with the cosmic ray trigger. The acquisition window spans over a period of 100~$\mathrm{\micro}$s, starting 20~$\mathrm{\micro}$s before the trigger and ending 80~$\mathrm{\micro}$s after it. Each waveform is made of 100000 bins (also called samples) with 1~ns time difference between.  Due to the drift velocity of 0.28~cm/$\mathrm{\micro}$s, a signal is expected to occur within 0 and 40~$\mathrm{\micro}$s drift time. An additional period of 40~$\mathrm{\micro}$s is acquired in order to study the response of the TPC up to 80~$\mathrm{\micro}$s after the trigger. Figure~\ref{fig:WF_Exp} shows a signal centred around  
 7~$\mathrm{\micro}$s, meaning that the primary electrons are created  roughly 2~cm upstream of the ThGEM.



The PMT baseline shows a very low frequency modulation. To take this into account, we develop a baseline subtraction algorithm based on a moving average of 2~$\mathrm{\micro}$s, corresponding to 2000 samples. First, the RMS and the average of the first 2000 sampled ADC values are calculated, and the next sample (2001$^{\text{st}}$) is shifted by this average. Following this, it is determined whether sample 2001 is higher or lower than 4 times the RMS determined earlier. 
\begin{itemize}
    \item If it is lower, a new RMS and average are determined for samples 2 to 2001, and sample 2002 is shifted by this updated average.
    \item If it is higher, it means that the sample contains signal, and therefore the baseline shift value remains unchanged, and sample 2002 is shifted by the initial average. 
\end{itemize} 
This procedure is applied to all the samples of the waveform, providing a simple but effective algorithm for the treatment of the baseline shift.

\subsection{Photomultiplier calibration}
\begin{figure}[htb]
    \centering
    \includegraphics[width = 0.5\textwidth]{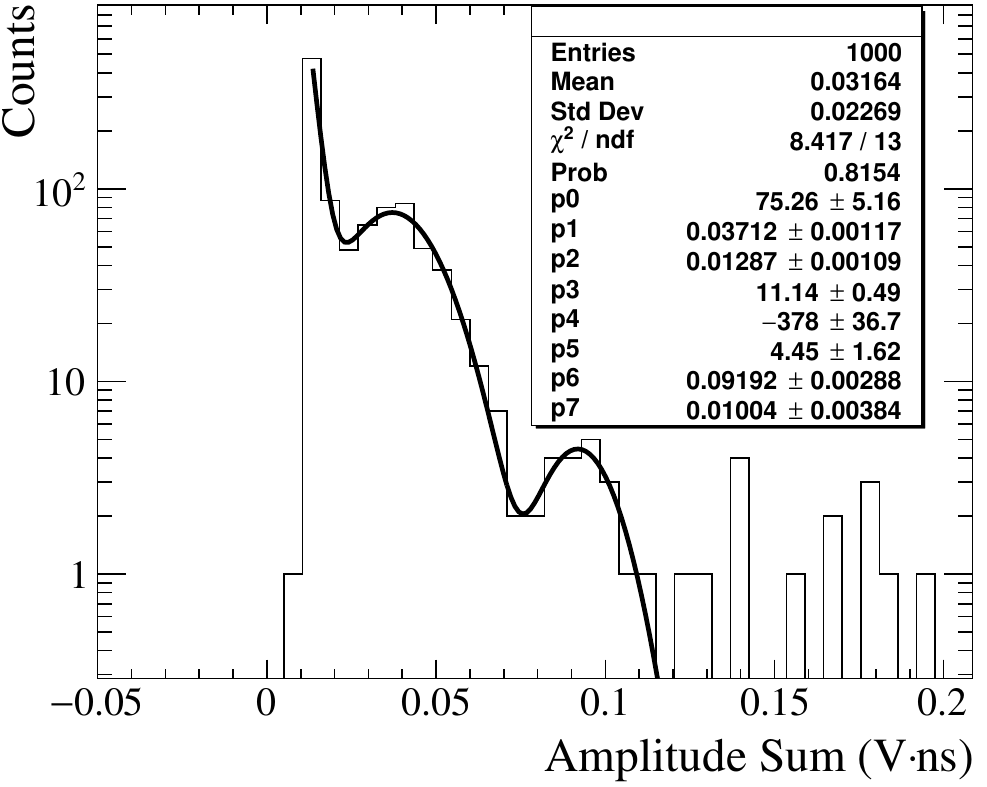}
    \caption{Amplitude sum distribution of the PMT's dark noise. A specific noise peak can be seen at low values, followed by two peaks corresponding to one, respectively two photo-electron pulses. The distribution is fitted with a function obtained through the sum of an exponential and two Gaussians. }
    \label{fig:PMT_Calibr}
\end{figure}

The calibration of the PMT relies on spontaneously emitted single thermal electrons within the PMT photo-cathode.  These electrons are multiplied between the PMT dynodes in the same manner as photo-electrons, and therefore generate the same output signal. 
The data used for PMT calibration are taken with a blinded PMT and a sampling frequency of 1~GHz. In these data, we anticipate the presence of electronic noise along with small polarised signals caused by the emission of thermal electrons. In our analysis, we want to isolate pulses induced by one or two thermal electrons. To do this, the dataset is analysed sample by sample. A thermal electron pulse candidate is defined as a group of one, or at most two consecutive samples, all exceeding the threshold of 4~RMS. These pulses are stored and used for calibration.
 The approximate integral (amplitude sum) of such pulses is determined by adding the areas of the pulse samples. Figure~\ref{fig:PMT_Calibr} shows the distribution of the amplitude sum throughout the calibration data taking. This shows a prominent peak towards low values and two smaller peaks specific to one and two photons. The distribution is well fitted by an exponential function summed to two Gaussians, yielding mean values for one-photon and two-photon pulses at $\mathrm{\micro}_\text{1}$ = 0.037 $\pm$ 0.001~V$\cdot$ns, respectively, $\mathrm{\micro}_\text{2}$ = 0.092 $\pm$ 0.002~V$\cdot$ns. $\mathrm{\micro}_\text{2}$ is slightly larger than the expected value of 2$\cdot\mathrm{\micro}_\text{1}$ as sensitivity is lost in the integral calculation for smaller pulses.

The nominal calibration value provided by Hamamatsu ( 0.08 V$\cdot$ns ) is  higher than our estimate of  $\mathrm{\micro}_\text{1}$ = 0.037 $\pm$ 0.001~V$\cdot$ns. Both values are rough estimates, and the very way in which we calculate the integral implies that our estimate is a slight underestimate because it does not include the integral contribution of the falling/rising edge-bin of the analysed pulses. This disagreement is not important for this paper as the conclusion remains unchanged no matter what calibration value we use. 
The value of $\mathrm{\micro}_\text{1}$ = 0.037 $\pm$ 0.001~V$\cdot$ns is going to be used in the following sections as calibration value in order to determine the number of photo-electrons (simply named photons hereafter) corresponding for any pulse shape present on the waveform.

\section{Results}
\label{sec:Results}

\subsection{Waveform sum }

\begin{figure}[htb]
         \centering
         \includegraphics[width=0.7\textwidth]{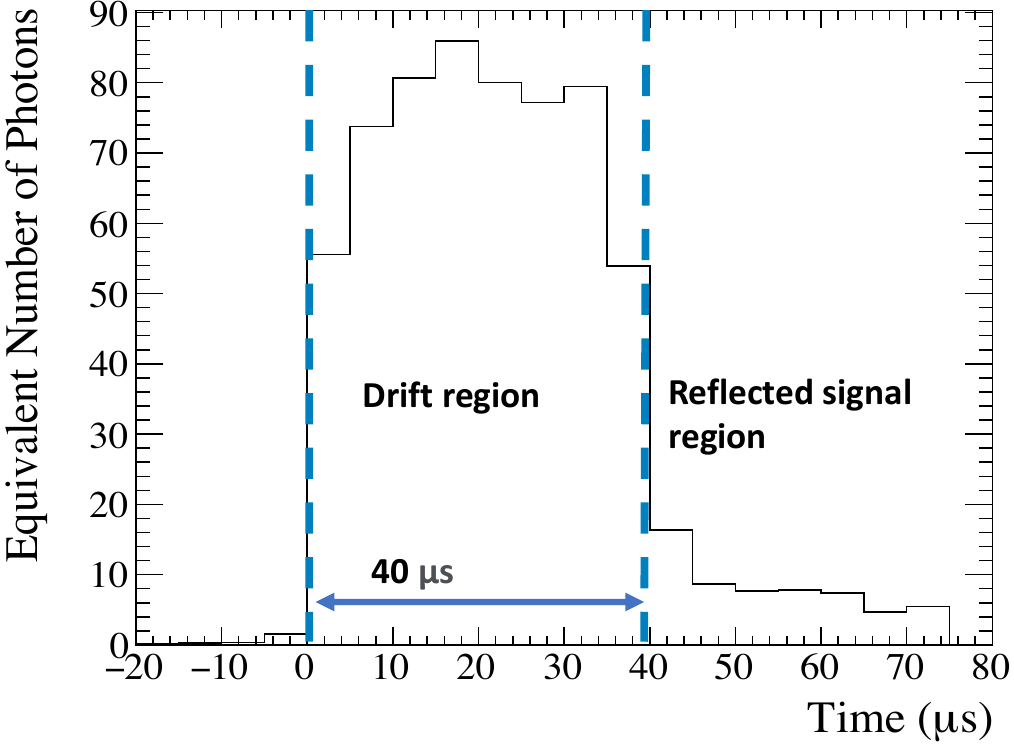}
         \caption{Waveform sum for several thousand signal events: a large number of photons is observed for drift times below the estimated maximum drift time of 40~$\micro$s, with a tail in the distribution between 40~$\mathrm{\micro}$s and 80~$\mathrm{\micro}$s. This arises from electrons extracted from the cathode, which travel the whole drift length.  }
         \label{fig:WF_Sum_1}
\end{figure}

In order to prove the functionality of the chamber, a first data set is taken with the standard drift settings outlined in Table~\ref{tab:OperatingConditions} and with no voltage on the mesh. 
To reject empty waveforms due to triggers from particles not crossing the drift volume, only events containing at least one sample with amplitude 4 times above the baseline RMS are selected as candidate TPC signals. The waveforms of several thousand signal events are summed together; see figure~\ref{fig:WF_Sum_1}. An accumulation of entries is visible in the drift region (from 0  to 40 $\micro$s) proving the correct operation of the TPC. To emphasise the contribution due to photons, only samples compatible with signal (above pedestal threshold) are included in the Waveform Sum (WFS). Following the waveform sum, the number of entries in each bin is divided by the total number of signal events. The calibration described earlier is used to convert the vertical axis to the expected number of photons. Therefore, the height of each bin gives the average number of photons detected per signal event during the time window described by the bin width.

The width of the sum is approximately 40~$\mathrm{\micro}$s, which is in agreement with the expected value based on the electron drift velocity in pure argon. The drift velocity as a function of the electric field is predicted using the Garfield++~\cite{Garfield} framework of the gas simulation program MAGBOLTZ~\cite{Magboltz}.

The WFS shows a sharp rise at 0~$\mathrm{\micro}$s, which is almost synchronous with the cosmic ray trigger telescope. The shape of the WFS during the drift period shows a slight increase at the beginning, which might be caused by the non-uniformity induced by the cosmic ray trigger geometry. Furthermore, the shape of the waveform sum remains high almost for the entire drift region; a good indication of excellent gas purity.

\begin{figure}
     \centering
     \begin{subfigure}[b]{.45\textwidth}
         \centering
        \includegraphics[width = \textwidth ]{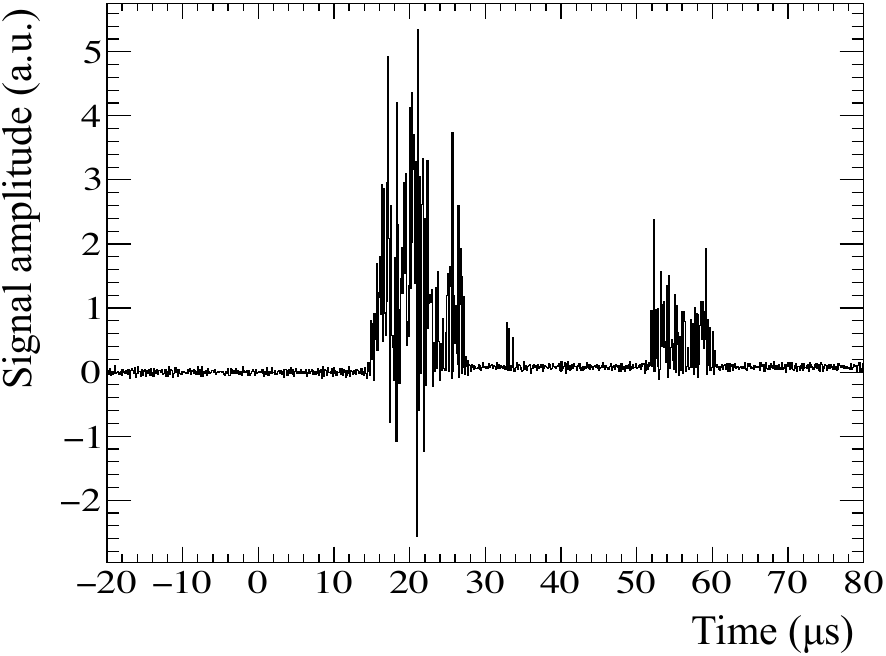}
         \caption{}
         \label{fig:WF_example}
     \end{subfigure}
     \hfill
     \begin{subfigure}[b]{0.45\textwidth}
         \centering
        \includegraphics[width = \textwidth]{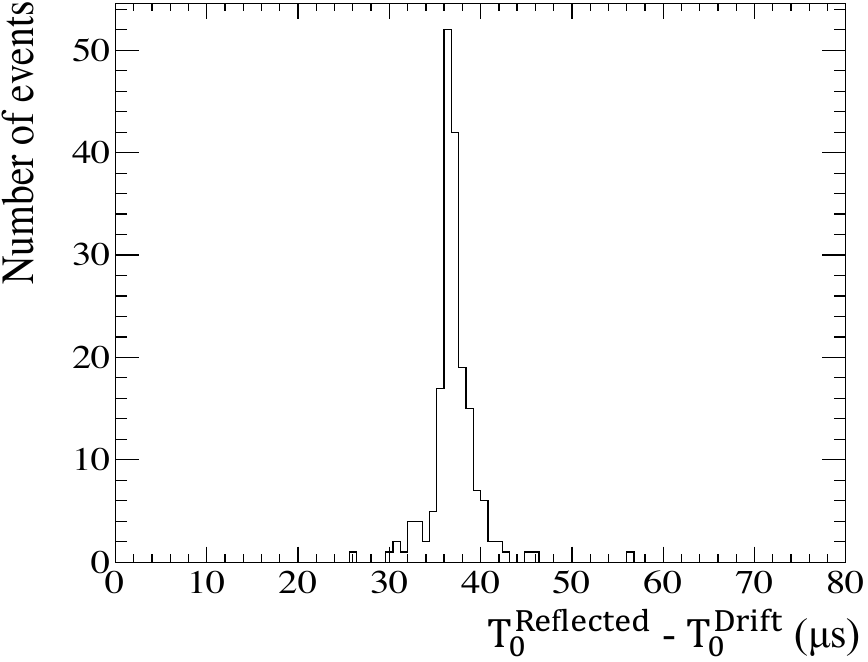}                 
         \caption{}
         \label{fig:TimeDifference}
     \end{subfigure}
        \caption{(a) Waveform example with a strong signal in the drift region ($\sim$ 20~$\micro$s) and its reflection at $\sim$ 55~$\micro$s. (b) Distribution of time difference between the first photon arrival during the drift period and the first photon arrival for the reflected period. The difference peaks at $37$~$\micro$s. }
\end{figure}

Beyond the $40$~$\mathrm{\micro}$s window the WFS decreases rapidly and a tail extending up to 80 $\micro$s is visible.  This tail occurs because some of the VUV photons produced in the ThGEM strike the copper cathode, extracting electrons via the photoelectric effect. These electrons drift towards the ThGEM, traversing the whole drift volume and therefore, generate a signal about $40$~$\mathrm{\micro}$s after the primary signal. We checked this hypothesis by plotting, event by event, the arrival time difference between the photons observed in the drift period and those in the reflected signal region. Figure~\ref{fig:WF_example} shows an event where the direct and the reflected signal are visible. The direct signal shows a strong component in the negative axis, given by the restoration of the PMT following a large positive signal. This does not impact the results, as the waveform integral is not used to determine the time difference between the direct and reflected signal, shown in figure~\ref{fig:TimeDifference}. To avoid the leakage of photons due to electron diffusion and electric field inhomogeneity, we select a time period in the centre of the drift time and the reflected signal region. The difference peaks at 37~$\mathrm{\micro}$s, which is around the drift time from the cathode, as expected from the Garfield~\cite{Garfield} prediction of the drift velocity (0.28~cm/$\mathrm{\micro}$s, yielding a total drift time from the cathode of 41~$\mathrm{\micro}$s). This favours the hypothesis of electrons extracted from the cathode by the photoelectric effect.  

 \subsection{Enhanced light yield }

\begin{figure}
    \centering
    \includegraphics[width = 0.7\textwidth , height = 0.5\textwidth]{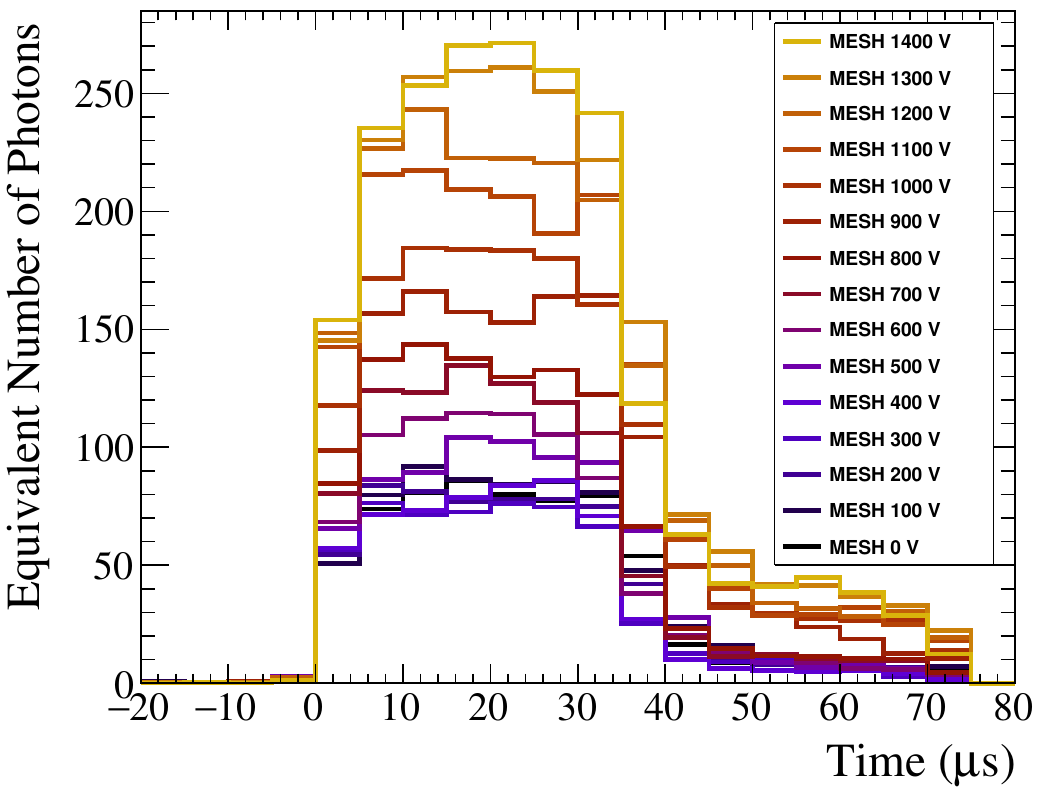}
    \caption{Waveform sum stack for several voltages applied on the mesh. It can be observed that generally, the number of photons detected increases with the voltage applied and that the photon yield does not vary significantly for mesh voltages below 500~V. 
    } 
    \label{fig:WF_Sum_Stack}
\end{figure}

In the previous section it was established that light produced by the ThGEM is enough to detect cosmic rays using the present TPC. In order to enhance further the light yield, a voltage is applied to the mesh and is gradually increased.
For each of these voltages, the waveform sum is plotted in figure~\ref{fig:WF_Sum_Stack} similarly to the previous section.  It is observed that the waveform sum is similar for mesh voltages up to 400~V,  while above that, it increases in amplitude with increasing mesh voltage.

\begin{figure}[htb]
    \centering
    \includegraphics[width = 0.7\textwidth]{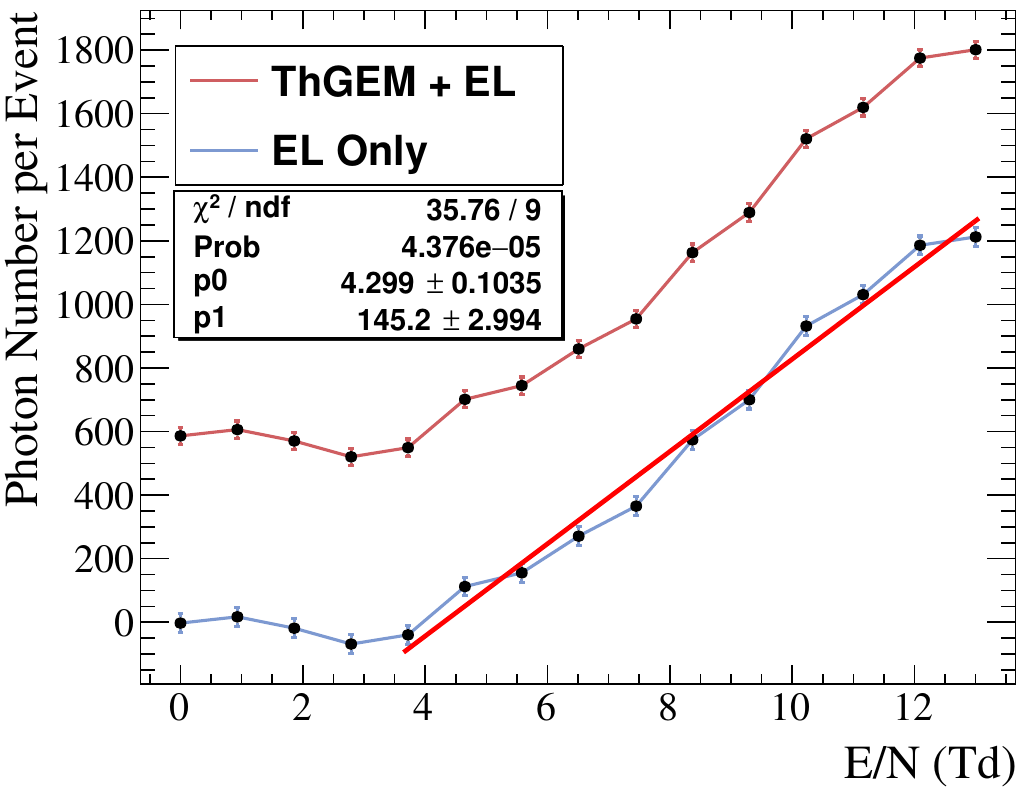}
    \caption{Number of photons as function of the mesh voltage. The red line represent the light yield from both the ThGEM and EL. The blue line represents the isolated contribution of the EL only. }
    \label{fig:AbsGain}
\end{figure}

For any voltage applied to the mesh, the shape of the histogram maintains the same structure, in which light contributions from both the drift and reflected regions are visible. We will include both in the analysis to explore the electroluminescence enhancement, considering that both signals go through the amplification process. In order to quantify this increase, it is enough to take the sum of bins of the waveform sum during the expected drift time, which gives the average number of photons detected per signal event. In figure~\ref{fig:AbsGain}, the number of photons is plotted against the reduced electric field (E/N) within the EL gap. E/N is correlated to the mesh voltage and the gas parameters  within the chamber and is expressed in units of Townsend: 1~ Td = 10$^{-17}$ V~cm$^2$~/~atom.

The number of photons is approximately constant up to a reduced field of $\sim$ 4~Td. This nonzero contribution is given by the ThGEM, across which the voltage drop is constant throughout the complete data taking. As the electric field within the ThGEM is much higher than the EL gap electric field, the ThGEM photon yield is not expected to change with different mesh voltages. 
The number of photons produced in the ThGEM mainly depends on the potential applied between the two ThGEM plates, and it is measured to be  $N_{\gamma}^{ThGEM} = 589 \pm 24$ photons per signal event. However, there is a slight dip in the photon yield between 2~Td and 4~Td. This is because data were acquired over several weeks, and there might have been small changes in the setup that were out of our control. Such systematic uncertainties are not included in the analysis, leading to an overall underestimation of the errors. 


The EL contribution is isolated by offsetting all the points on the red line (figure \ref{fig:AbsGain}) by the ThGEM contribution, $N_{\gamma}^{ThGEM}$. The obtained data points (light blue) are fitted with a linear function $p_1(E/N-p_0)$, where $p_0$ represents the minimum excitation energy for argon, and $p_1$ is related to the photon yield per electron as explained later. Overall, the total light yield captured by the PMT can be quantified by:

\begin{equation}\label{eq:AbsGain}
    Y = N_{\gamma}^{ThGEM} + \alpha_{EL} N_{e^-}^{ThGEM}, 
\end{equation}
where  $N_{\gamma}^{ThGEM}$ represents the number of photons originating from the ThGEM,  $N_{e^-}^{ThGEM}$ is the number of secondary electrons produced in the ThGEM holes, and $\alpha_{EL}$ is the photon yield per electron due to the electroluminescence effect. Since we do not know the number of electrons produced in the avalanche, we rewrite the equation as: 

\begin{equation}
    Y = (1  + \alpha_{EL} f_{\gamma,e-} ) N_{\gamma}^{ThGEM},
\end{equation}
where $f_{\gamma,e-} = \frac{N_{e^-}^{ThGEM}}{N_{\gamma}^{ThGEM}}$ is the ratio between the flux of $\gamma$'s and electrons produced in the ThGEM avalanche\footnote{There is a geometrical acceptance difference between photons produced in the avalanche and those produced during the transport along the EL region that we ignore to simplify the discussion.}.

The number of photons produced in the EL gap per electron is well described by the phenomenological linear dependency~\cite{Monteiro:2008zz}:

\begin{equation} 
    \alpha_{EL}=\begin{cases}
    0&E/N\leq E_0/N\\
    \beta \left( E/N - E_0/N \right)&E/N> E_0/N,
    \end{cases}
\end{equation} 
where $E/N$ is the reduced field across the EL gap. $E_0/N$ provides an estimation of the excitation threshold for argon, and $\beta$ is the coefficient of linear dependency with the applied reduced electric field difference $\left( E/N - E_0/N \right)$. Combining all the previous equations we obtain the electroluminescence contribution: 


\begin{equation}\label{eq:AbsGainRenorm}
   \alpha_{EL} f_{\gamma,e-}  N_{\gamma}^{ThGEM} =\begin{cases}
    0&E/N\leq E_0/N\\
    \beta \left( E/N - E_0/N \right) f_{\gamma,e-}  N_{\gamma}^{ThGEM} = p_1 ( E/N - p_0) &E/N> E_0/N,
    \end{cases}
\end{equation}
with $p_0$ being the excitation threshold for argon ($E_0/N$)  and $p_1$ the linear coefficient multiplied to the number of electrons after the ThGEM amplification ($\beta f_{\gamma,e-}  N_{\gamma}^{ThGEM} = \beta  N_{e-}^{ThGEM} $). A linear fit to the number of photons as function of the reduced field, see figure~\ref{fig:AbsGain}, yields to a value of $p_1$ equal to  $145 \pm 3$~photons/Td.
The crossing point of the linear fit with zero ($p_0$) determines the minimum field required to initiate the electroluminescence at $p_0$ = $4.3\pm0.1$~Td. This threshold is in good agreement with results obtained under different experimental conditions: 4.1~$\pm$0.1~Td in~\cite{OLIVEIRA}, $\sim$~4~Td in~\cite{ArgonLight:BUZULUTSKOV}. 
We derive the value of $\beta$ from equation~7 in \cite{Monteiro:2008zz} to be $8.10$~photons/electron/Td after correcting for the different EL gap width and pressure in the experimental setups. Since $\beta$ is computed for a 4$\pi$ acceptance we need to divide by 2 to accommodate to our forward geometry acceptance. An additional 10\% reduction is obtained from a geometrical simulation of our finite setup.  The $\beta$ value and the estimations of $p_1$ and $N_{\gamma}^{ThGEM}$ obtained in this study (see figure~\ref{fig:AbsGain}) allow us to estimate the value of the fraction of photons detected versus electrons in the ThGEM (1/$f_{\gamma,e^-}$) to be $18.3$~photons/electron. This method could be utilised, upon proper calibration, to study the ThGEM gain separated for photons and electrons.  

Throughout the measurements, the mesh is stable and the reduced electric field represents the threshold from where electron amplification would occur in the EL gap. Operating the mesh at a voltage of $1400$~V($13$~Td), $1801 \pm 27$ photons per signal event are measured. Comparing this value to the ThGEM-only photon yield ($N_{\gamma}^{ThGEM} = 589 \pm 24$~photons per signal event), 3 times more photons are measured. This increase in light yield is only due to EL photons produced in the EL gap, meaning that more light can be obtained without a negative impact on the energy resolution. However, with the mesh, an additional distance is placed between the ThGEM and the PEN, and photons are further dispersed. Should this technology be used for tracking, the distance before mentioned would negatively impact the position resolution, and for such reasons the EL gap length has to be optimised. In our current setup, it is assumed that the EL gap of 0.44~cm would not significantly impact the position resolution as the electron diffusion in pure argon gas dominates the photon smearing between the ThGEM and PEN. To avoid instability, values of the reduced electric field larger than 13~Td could not be explored. Increasing the EL gap electric field beyond this point causes numerous sparks that make the high voltage setup unstable. The sparks are probably due to the presence of defects on the mesh, causing non-negligible charge displacement.

\subsection{Comparison with predictions }

\begin{figure}
     \centering
     \begin{subfigure}[b]{.45\textwidth}
         \centering
        \includegraphics[width = \textwidth ]{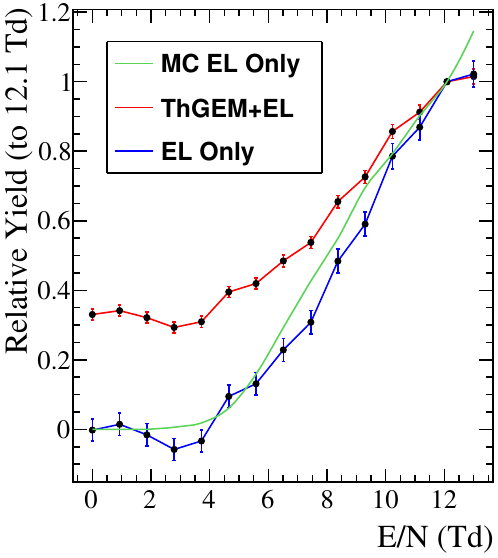}
         \caption{}
         \label{fig:RelYield_a}
     \end{subfigure}
     \hfill
     \begin{subfigure}[b]{0.45\textwidth}
         \centering
        \includegraphics[width = \textwidth ]{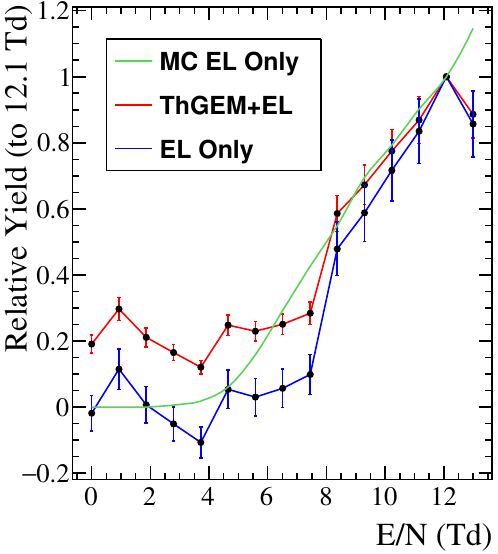}                 
         \caption{}
        \label{fig:RelYield_b}  
     \end{subfigure}
        \caption{ Relative yield as function of the EL gap reduced electric field. The yield is determined for the drift period (a) and after the drift period (b). The normalisation is done to 12.1~Td, point common to both our data sets and the model from \cite{OLIVEIRA}. }    
        \label{fig:MC_DataComp}
\end{figure}

Comparisons between different experiments are notoriously difficult due to differences in geometries and several other properties which are hard to control. Nevertheless, we can numerically check the current results using the theoretical prediction from~\cite{OLIVEIRA}. The blue line from  figure~\ref{fig:AbsGain} shows the average number of EL photons detected per event, while the theoretical prediction from \cite{OLIVEIRA} gives the reduced electroluminescence yield (in units of photons electron$^{-1}$ cm$^2$ atom$^{-1}$). In order to compare them, a new quantity, the relative yield, is defined:  

\begin{equation} \label{eq:rel_y}
 Y_{Rel} = \frac{Y_{i}}{Y_{12}},
\end{equation}where $Y_i$ is the photon number at a specific reduced electric field, $i$, and $Y_{12}$ is the photon number at 12.1~Td. $Y_{Rel}$ is calculated for both the theoretical model and our data set (figure~\ref{fig:AbsGain} - blue line) and plotted in figure~\ref{fig:MC_DataComp}. On the left-hand side, the plot shows the relative yield against the reduced field in the EL gap for photons detected during the drift period. The theoretical model (green) predicts well the data points (blue), confirming the observation of electroluminescence light. 

A similar calculation is done for photons from the tails (between 40~$\mathrm{\micro}$s and 80~$\mathrm{\micro}$s), plotted in figure~\ref{fig:RelYield_b}. Here, the photon yield shows a generally increasing trend, motivated as follows. The number of photo-electrons extracted from the cathode is constant, produced by the ThGEM photons, without any dependence on the mesh voltage. The extracted electrons travel through the drift volume, get amplified in the ThGEM and propagate through the EL gap. As for the primary electrons, the amount of light produced by the cathode electrons propagating through the EL gap is linear with the EL gap reduced electric field. Compared to figure~\ref{fig:RelYield_a}, the points between 0~Td and 4~Td show a similar dip, and the points above 8~Td show a similar increase. The only difference is observed between 4~Td and 8~Td, where the relative points do not increase as fast as in figure~\ref{fig:RelYield_a}. Due to the low photon statistics in this region, the separation of the background is not simple to achieve. Therefore we expect many issues when the photon yield is low, but when the photon gain is large, the background separation is restored. This is why we only trust values above 8~Td.
To conclude on the tail photons, the overall increasing trend in figure~\ref{fig:RelYield_b} favours our hypothesis that the tails originate from photoelectric electrons extracted from the cathode copper plate. 


 For completeness, the total relative yield (ThGEM + EL) is included in figures~\ref{fig:RelYield_a} and~\ref{fig:RelYield_b}. This is obtained similarly as in equation~\ref{eq:rel_y} but this time using the total photon yield shown in figure~\ref{fig:AbsGain} (red line). Due to the constant, non-zero contribution of the ThGEM, for both figures~\ref{fig:RelYield_a} and~\ref{fig:RelYield_b}, the total photon shows a clear deviation from the EL prediction at low electric fields. As the EL gap photon contribution becomes important, the total relative yield converges towards the trend of the EL prediction.

To conclude on this subsection, the generally increasing trends observed in both figures~\ref{fig:RelYield_a} and \ref{fig:RelYield_b} are a further confirmation of the fact that the light detected with this TPC has a big contribution from the EL gap. This confirms the physics of the EL gap technology and paves the way towards new studies.



\section{ Conclusion and future prospects}
\label{sec:Concl}
We presented the construction and operation of an optical gaseous TPC, with light production based on a ThGEM and an additional region of low electric field called the EL gap. The TPC is read by a photomultiplier tube and uses PEN as a wavelength-shifting material. We have shown that a combined ThGEM and EL gap electric field can increase significantly, up to three times the photon yield. This technique can be of interest for many applications where a low number of photons is expected. Most notably, it can make the difference in whether a particle of low energy and low primary ionisation can be detected or not.

Other than photon enhancement, the EL gap technology can be used to improve the energy resolution in general. A promising application is for detectors based on ThGEMs where a sufficient quantity of photons is expected. In such cases, the ThGEM gain could be decreased, and the EL gap field could be increased such that there is no net change in the observed photon yield. Having less charge amplification in the ThGEM but the same number of photons could mean an improvement of the overall energy resolution.   

One of the main limitations of this paper is that the gain of the ThGEM is not measured directly. For future studies, the TPC is going to be upgraded with a charge measurement device coupled to the ThGEM down electrode and the mesh. This would play a key role in directly confirming this paper's results but also for resolution studies with radioactive sources such as $^{55}$Fe.

The photon amplification obtained in this study can be further improved by increasing the EL gap reduced field or the gap between the ThGEM and the PEN Mesh. The first option is not of great interest as the idea of the EL gap electric field is to produce light without charge amplification. The second option could be attractive; however, since increasing the gap would also decrease the tracking resolution, a balance needs to be found depending on the requirements of the specific application. 

The next step for the TPC is to implement this technology with a tracking device and increase the gas density, using the advantages of EL amplification to optimise photon yield and energy resolution. A Multi Pixel Photon Counter (MPPC) array has already been commissioned at the University of Geneva, and the PMT has been replaced with it. Currently, measurements are being taken with the upgraded TPC, and the possibility of full tracking is being investigated. As seen in this study, with a PMT of 2~cm diameter, enough light can be collected for cosmic ray detection. However, as the readout becomes more pixelated, as in the case of the MPPC array, the photon number interacting with each channel drastically decreases. The maximum pixelation of an MPPC array is therefore limited by the photon yield. Therefore, EL gap light enhancement can be used to allow for finer channel distribution. 

On a different base, the method described in this paper could be also used in dedicated setups to study the ThGEM gain separated for photons and electrons. 



   


\acknowledgments

This material is based upon work primarily supported by the Swiss National Science Foundation (SNSF) under research grant  Grant No. 200 021\_85012.  Assistance provided by the technical and engineering support from the University of Geneva (F.Cadoux, Y.Favre and D.Ferrer) and INFN Bari was greatly appreciated. T. Lux acknowledges funding from the Spanish Ministerio de Econom\'{\i}a y Competitividad (SEIDI-MINECO) under Grants No. PID2019-107564GB-I00. IFAE is partially funded by the CERCA program of the Generalitat de Catalunya. E. Roe acknowledges the support of the Boston University through the Boston University Geneva exchange program.

\bibliographystyle{unsrt}
\bibliography{references.bib}

\end{document}